\documentstyle[psfig,aps,pre,twocolumn,floats]{revtex}
\begin{document}
\newcommand{\drm}{ {\rm d}}

\wideabs{ {\small Eur. Phys. J. B {\bf 14}, 293-300 (2000)}\\
\bibliographystyle{prsty}
\title{Quasiclassical Hamiltonians for large-spin systems}
\author{
D. A. Garanin \cite{e-gar}, K. Kladko \cite{e-kla}, and P. Fulde$^\ddagger$
}

\address{{\normalsize$^{*,\ddagger}$}Max-Planck-Institut f\"ur Physik komplexer
Systeme, N\"othnitzer
Strasse 38,
D-01187 Dresden, Germany\\ }

\address{{\normalsize$^\dagger$}Condensed Matter Physics Group and Center
for
Nonlinear Studies,
Theoretical Division, LANL, MS-B258, Los Alamos, NM 87545, USA \\
}

\smallskip
\date{Received 5 May 1999, revised 24 September 1999}

\maketitle

\begin{abstract}
We extend and apply a previously developed method for a semiclassical treatment  of a
system with large spin $S$.
A multisite Heisenberg Hamiltonian is transformed into an effective classical
Hamilton function which can be treated by standard methods for classical systems.
Quantum effects enter in form of {\em multispin} interactions in the Hamilton
function.
The latter is written in the form of an expansion in powers of $J/(TS)$, where $J$ is the
coupling constant.
Main ingredients of our method are spin coherent states and cumulants.
Rules and diagrams are derived for computing cumulants of groups of operators
entering the Hamiltonian.
The theory is illustrated by calculating the quantum corrections to the free energy of a
Heisenberg chain which were previously computed by a Wigner-Kirkwood expansion.
\end{abstract}
\smallskip
\begin{flushleft}
\end{flushleft}
}

\vspace{-1cm}

\section{Introduction}

Quantum effects in systems with localized spins depend on the size of the spin $S$.
They are strongest for $S=1/2$ while systems behave classically in the limit
$S\to\infty$.
There are, however, numerous compounds having intermediate or high spin values,
which we would like to treat semiclassically, i.e., classically with quantum corrections
expanded in powers of $1/S$.
Different proposals have been made in the past how this can be achieved \cite{tsu84,ulyzas92}.
We have recently suggested a method by which we can derive effective classical
Hamilton functions for spin Hamiltonians by using spin coherent states and
cumulants \cite{klafulgar99epl}.
The quantum partition function is thereby replaced by a classical one, and the trace
over a complete set of quantum states is replaced by an integration ober the classical
spin vectors.
The mathematical structure of these two objects is quite different, and the transition
from one to the other was investigated by Lieb \cite{lie73} who obtained upper and lower
bounds for the quantum partition function for large $S$.
He did not consider the $1/S$ corrections to the classical partition function which are
discussed here.

The previous paper \cite{klafulgar99epl}  dealt with one- and two-spin systems and their semiclassical
description when $S$ is large.
Here we extend these calculations to a many-spin system.
The necessary formalism is described in Sec.\ \ref{basic}.
The method is applied in Sec.\ \ref{cumexp} to the Heisenberg Hamiltonian in an applied
magnetic field first stating the rules for evaluating the required cumulants.
As we will show this can be done by means of a diagrammatic technique.
The explicit form of the corresponding classical Hamilton function is derived in Sec. \
\ref{EffHam}.
All terms of order $J[J/(TS)]^2/S^n$ with $n=0,1,2$ are included in the effective
Hamilton function.
Obviously, the semiclassical description of large-spin systems breaks down in the
low-temperature limit, where quantization of energy levels becomes important.
To check the method, we use in Sec.\ \ref{chain} our effective Hamilton function in
order to calculate the free energy and other thermodynamic properties of a spin chain
to order $1/S^2$.
These results were previously computed by a Wigner-Kirkwood expansion without
using effective classical Hamiltonians.
Sec.\ \ref{discussion} contains a discussion and the conclusions.

\section{Spin coherent states and cumulants}
\label{basic}

To achieve our goal, we use two theoretical tools.
The first one is the coherent-state representation of quantum-statistical
averages.
On each lattice site $i$, we introduce  {\em spin  coherent states}
$|{\bf n}_i\rangle$, i.e., states with the maximal projection on the axis pointing in the direction of the
unit vector ${\bf n}_i$.
The direct product of these states $| \{ {\bf n}_i \} \rangle$ approaches the ``classical''
state of the spin system in the limit $S\to\infty$.
On the other hand, the basis of coherent states is complete (and even
overcomplete), so
that it contains all quantums states.
The quantum-statistical averages of the system can be rewritten in the
coherent-state
basis
with the help of the unity operator
%\marginpar{ResUnity}
%
\begin{equation}\label{ResUnity}
{\bf 1} = \frac{2S+1}{4 \pi} \int \drm{\bf n}\, |{\bf n}\rangle \langle{\bf n}|.
\end{equation}
For a single-spin system, the trace of an operator $\hat A$ over any complete
orthonormal
basis $|m\rangle$ becomes \cite{lie73}
%\marginpar{TraceRewrite}
%
\begin{eqnarray}\label{TraceRewrite}
&&
{\rm tr}\, \hat A = \sum_m \langle m| \hat A | m \rangle
= \frac{2S+1}{4 \pi} \int \drm {\bf n}
\sum_m \langle m|\hat A|{\bf n}\rangle \langle {\bf n}|m\rangle
\nonumber\\
&&\qquad\qquad
= \frac{2S+1}{4 \pi} \int \drm {\bf n}\, \langle {\bf n}| \hat A| {\bf n}\rangle.
\end{eqnarray}
Therefore the partition function for a many-spin quantum Hamiltonian $\hat H$ can be written as
%\marginpar{ZCohState}
%
\begin{equation}\label{ZCohState}
{\cal Z} =
\left(\frac{2S+1}{4 \pi} \right)^N \int \prod_{i=1}^N \drm {\bf n}_i\,
\langle \{ {\bf n}_i \}| \exp (- \beta \hat H) |
\{ {\bf n}_i \}\rangle,
\end{equation}
where $\beta \equiv 1/T$.
It has the same form as the one for {\em classical} systems, provided one defines the effective classical Hamilton function ${\cal H}$ by the relation
%\marginpar{HClass}
%
\begin{eqnarray}\label{HClass}
&&
\beta {\cal H} = - \ln \langle  \{ {\bf n}_i \}| \exp (- \beta \hat H) | \{ {\bf n}_i \}
\rangle
\nonumber \\
&&\qquad
{}=\langle \{ {\bf n}_i \}| 1 - \exp (- \beta \hat H) | \{ {\bf n}_i \} \rangle^c
\end{eqnarray}
Here the superscript $c$ denotes the cumulant of a matrix element (see below).
Cumulants constitute the second theoretical tool used in this paper.

The above formula is exact for any quantum spin system.
In the limit $S\to\infty$, ${\cal H}$ reduces to the usual classical Hamiltonian.
A mathematically rigorous treatment of this limiting transition can be found in
Lieb's paper, Ref.\ \cite{lie73}
Here we will obtain quantum corrections to the classical Hamiltonian using a
systematic
$1/S$ expansion for $S\gg 1$.
In the same context, a technique using partial differential equations for the density
matrix was applied in Ref.\ \cite{tsu84}.
But an explicit derivative-free expression for the effective classical
Hamiltonian was obtained in Ref. \cite{tsu84} only for the spin-chain model under
the condition that neigbouring spins are nearly collinear.
The method presented here does not use derivatives and yields
explicit results for any lattice.

The effective Hamilton function is obtained in form of a cumulant expansion by expanding in  Eq.\ (\ref{HClass}) the operator $\exp (- \beta \hat H)$.
It is worth noting a basic relation between semiclassical theory.
A characteristic property of quantum mechanics is that the expectation value of a product $A_1 A_2$ of two observables with respect to a quantum state $|\psi\rangle$ is generally distinct from the product of the expectation values of
$A_1$ and $A_2$.
The difference is just the cumulant, i.e.,
%\marginpar{QuantumNonZero}
%
\begin{equation}\label{QuantumNonZero}
\langle A_1 A_2\rangle^c = \langle A_1 A_2\rangle
- \langle A_1\rangle \langle A_2\rangle \ne 0.
\end{equation}
In the classical limit this difference vanishes.
Therefore we expect that a theory formulated in terms of cumulants is particularly suitable for a semiclassical expansion.

Let us recall the definition and basic properties of cumulants.
Averages or matrix elements $\langle \ldots\rangle$ can be expressed through
{\it cumulants} $\langle \ldots\rangle^c$ as follows
%\marginpar{AvrToCum}
%
\begin{eqnarray}\label{AvrToCum}
&&
\langle  A \rangle=\langle  A \rangle^{c},
\qquad
\langle  A_1 A_2  \rangle=\langle A_1 A_2  \rangle^{c}
+
\langle  A_1 \rangle \langle A_2  \rangle,
\nonumber\\
&&
\langle  A_1 A_2 A_3 \rangle = \langle  A_1 A_2 A_3 \rangle^{c}
+\langle  A_1 \rangle \langle A_2 A_3 \rangle^{c}
+\langle  A_2 \rangle \langle A_1 A_3 \rangle^{c}
\nonumber\\
&&\qquad\qquad
{}+\langle  A_3 \rangle \langle A_1 A_2 \rangle^{c}
+\langle  A_1 \rangle \langle  A_2 \rangle \langle  A_3 \rangle,
\end{eqnarray}
etc., where $A_i$ are classical stochastic variables or
quantum-mechanical operators.
The averaging above is performed over a classical distribution function or weighed
over
quantum states.
A more detailed discussion of cumulants can be found in refs.\
\cite{kub62,ful95book,klaful98}.
Cumulants can be obtained by differentiation of a generating function,
i.e., from
%\marginpar{AvrToCumLog}
%
\begin{equation}\label{AvrToCumLog}
\langle A_{1}^{k_1}\ldots A_{n}^{k_n} \rangle^{c}= \frac{\partial^{k_1}}{\partial
\lambda_1^{k_1} }\cdots \frac {\partial^{k_n}}{\partial
\lambda_n^{k_n}}~
\!\!\ln \langle  e^{\lambda_1 A_1} \ldots e^{\lambda_n A_n}
 \rangle
\end{equation}
at $\lambda_{1}=\ldots=\lambda_{n}=0$, in contrast to averages
$\langle\ldots\rangle$,
which are given by a
similar
expression without logarithm.
By multiplying by $\prod_{l=1}^n\lambda^{k_l}/k_l!$ and summing over all
$k_l=0,\ldots,\infty$ without the term $k_1=\ldots=k_n=0$ one obtains the identity
%\marginpar{ExpCumRel}
%
\begin{equation}\label{ExpCumRel}
\langle e^{\lambda_1 A_1}\ldots e^{\lambda_n A_n} -1 \rangle^c =
\ln\langle e^{\lambda_1 A_1} \ldots e^{\lambda_n A_n} \rangle.
\end{equation}
The second line of Eq.\ (\ref{HClass}) is a particular case of this formula.

Let us consider cumulants of spin operators with respect to spin coherent states.
These cumulants have an especially simple form when the spin-operator components
$S_z$ (axis $z$ along ${\bf n}$) and $S_\pm=S_x\pm iS_y$ are used.
A cumulant vanishes if it is  of the form $\langle \ldots S_+\rangle^c$,
$\langle \ldots S_+S_-\rangle^c$, $\langle \ldots S_z\rangle^c$, etc., where
$\ldots$
stands for any combination of spin operators.
That is, the number of $S_+$ and $S_-$ in the cumulant must be balanced in order to
give a
nonzero result.
If, however, this balance is achieved already within a subgroup of operators on the right or
left
side of the operator list, the cumulant vanishes, too.
Non-vanishing cumulants are
%\marginpar{LowestCums}
%
\begin{eqnarray}\label{LowestCums}
&&
\langle S_z \rangle = S, \qquad
\langle S_+ S_z^n S_- \rangle ^c = 2S (-1)^n,
\nonumber\\
&&
\langle S_+ S_z^n S_+ S_z^m S_- S_z^{n'} S_- \rangle ^c = -4S (-1)^{n+n'}
(-2)^m,
\nonumber\\
&&
\langle S_+ S_+ S_+ S_- S_- S_-\rangle ^c
= 3 \langle S_+ S_+ S_- S_+ S_- S_-\rangle ^c = 24S,
\end{eqnarray}
etc.
The above results can be obtained recurrently, using Eq.\ (\ref{AvrToCum}) and
the commutation
relations $[S_z,S_-]=-S_-$ and $[S_+,S_-]=2S_z$, which remain valid inside
cumulants.
For scaled spins
${\bf S}/S$, each nonvanishing cumulant containing $n$ spin operators scales like
$1/S^{n-1}$.

An expansion of ${\cal H}$ in Eq.\ (\ref{HClass}) in powers of $1/S$ is obtained from
the Taylor series
%\marginpar{TaylorCum}
%
\begin{eqnarray}\label{TaylorCum}
&&
 \beta {\cal H} =
  \langle 1 - \exp (- \beta\hat H) \rangle ^c
\nonumber\\
&&
 =
  \beta\langle \hat H\rangle^c
 - \frac{\beta^2}{2!} \langle \hat H \hat H\rangle^c
 + \frac{\beta^3}{3!} \langle \hat H \hat H \hat H \rangle^c + \ldots,
\end{eqnarray}
where we have introduced the shorthand notation
$\langle\ldots\rangle \equiv \langle \{ {\bf n}_i \}| \ldots | \{ {\bf n}_i \} \rangle$.
Here the first term on the right-hand side is the classical energy of the spin.
As seen from Eq.\ (\ref{LowestCums}), increasing powers of $1/S$ appear
in each order of the expansion.
In fact, Eq.\ (\ref{TaylorCum}) is an expansion in powers of $J/(TS)$ and it breaks
down at
low
temperatures.
For $S\gg 1$, the  range of convergence is much larger than that of the high-temperature series expansion.
For the calculation of thermodynamic quantities one can further expand $\exp(-\beta
{\cal H})$ in
powers of $1/S$.
In that case the {\em statistical weights} of different spin orientations are
described by
a purely classical Hamiltonian, whereas quantum effects  manifest themselves in
corrections to the {\em density of states}.

\section{The cumulant expansion}
\label{cumexp}

Let us consider a spin Hamiltonian of the Heisenberg form
%\marginpar{QHam}
%
\begin{equation}\label{QHam}
\hat H = - \sum_i {\bf H}_i \cdot {\bf S}_i - \frac 12 \sum_{ij} J_{ij} {\bf S}_i \cdot
{\bf
S}_j .
\end{equation}
To implement the cumulant expansion of Eq.\ (\ref{TaylorCum}), it is convenient
to
express
the spin operator on each site $i$ in the coordinate system with the $z$ axis
along the
coherent-state vector ${\bf n}_i\equiv {\bf n}_{iz}$
%\marginpar{SpinCoord}
%
\begin{equation}\label{SpinCoord}
{\bf S}_i = \sum_{\alpha_i=z,\pm} {\bf n}_{i\alpha_i} S_{i\alpha_i}
    \qquad {\bf n}_\pm \equiv ({\bf n}_x \mp i{\bf n}_y)/2,
\end{equation}
where ${\bf n}_x$ and ${\bf n}_y$ are appropriate transverse basis vectors.
Insertion into Eq.\ (\ref{TaylorCum}) leads to expressions of the type
%\marginpar{Tay2}
%
\begin{eqnarray}\label{Tay2}
&&
- \frac{ \beta^2 }{ 2!\, 2^2 } \sum_{ii'jj'} J_{ii'} J_{jj'}
\sum_{\alpha_i \alpha_{i'} \alpha_j \alpha_{j'}}
({\bf n}_{i\alpha_i} \cdot {\bf n}_{i'\alpha_{i'}})
({\bf n}_{j\alpha_j} \cdot {\bf n}_{j'\alpha_{j'}})
\nonumber\\
&&\qquad\qquad
{}\times\langle (S_{i\alpha_i} S_{i'\alpha_{i'}}) (S_{j\alpha_j} S_{j'\alpha_{j'}})
\rangle^c,
\end{eqnarray}
as illustrated by the second-order pure-exchange term.
The brackets inside the cumulant imply that this cumulant is defined with respect
to the
{\em
two pairs} of the spin-operator components and not with respect to four single
operators.
To calculate such cumulants of composite operators, it is convenient to (i) express them
through
ordinary matrix
elements
[see, e.g., Eq.\ (\ref{QuantumNonZero})] and (ii) express matrix elements through
cumulants of
single spin-operator components [see Eqs.\ (\ref{AvrToCum})].
The latter are nonzero only if all the operators in the cumulant belong to the
same site
in which case
they are readily given by Eqs.\ (\ref{LowestCums}).
One can see that in several lowest orders of the cumulant expansion,
the summation over the spin-component indices $\alpha$ in Eq.\ (\ref{Tay2})
reduces to a
single
realization which gives a nonzero result.
The result of such a procedure applied to Eq.\ (\ref{Tay2}) is the following:
%\marginpar{CumToCum2}
%
\begin{eqnarray}\label{CumToCum2}
&&
\frac{ 1 }{ 2^2 } \langle (A_i A_{i'}) (A_j A_{j'}) \rangle^c =
\langle A_i \rangle \langle A_{i'} A_j \rangle^c \langle A_{j'}  \rangle
\nonumber\\
&&\qquad\qquad
{} + \frac 12 \langle A_i A_j \rangle^c \langle A_{i'} A_{j'} \rangle^c,
\end{eqnarray}
where $A_i\equiv S_{i\alpha_i}$.
Hereby we have taken into account that operators $A_i$ and $A_{i'}$ belong to different sites so that terms of the type $\langle A_i A_{i'} \rangle^c$ vanish.
On the rhs of this expression, one could add terms differing by permutations of the
indices $i, i'$ and/or $j, j'$.
These terms make the same contributions to Eq.\ (\ref{Tay2}) as those present in
Eq.\ (\ref{CumToCum2}); Instead of writing them explicitly, they lead to the prefactor
$1/2^2$ in Eq.\ (\ref{CumToCum2}).
Generally, the possibility of permutating spin operators connected by the
exchange
interaction
effectively cancels the coefficient 1/2 in Eq.\ (\ref{QHam}).
The factor 1/2 in front of the last term of Eq.\ (\ref{CumToCum2}) appears because
of the
permutation
of both $i, i'$ and $j, j'$ which does not generate a new term of this type.
For the cumulant with three groups of spin operators one obtains
%\marginpar{CumToCum3}
%
\begin{eqnarray}\label{CumToCum3}
&&
\frac{ 1 }{ 2^3 } \langle (A_i A_{i'}) (A_j A_{j'})  (A_l A_{l'})\rangle^c =
\nonumber\\
&&
%\big[
\langle A_i \rangle  \langle A_{i'} A_j \rangle^c  \langle A_{j'} A_{l} \rangle^c
\langle A_{l'}
 \rangle
\nonumber\\
&&
{}+ \langle A_i \rangle  \langle A_{i'} A_l \rangle^c  \langle A_{j} A_{l'} \rangle^c
\langle
A_{j'}  \rangle
\nonumber\\
&&
{} + \langle A_j \rangle  \langle A_i A_{j'} \rangle^c  \langle A_{i'} A_{l} \rangle^c
\langle
A_{l'}  \rangle
%\big]
\nonumber\\
&&
{} + \langle A_i A_j A_l \rangle^c \langle A_{i'} \rangle  \langle A_{j'} \rangle
\langle A_{l'}
\rangle
\nonumber\\
&&
{} + %\big[
\langle A_i A_j \rangle^c \langle A_{i'} A_{j'} A_l\rangle^c \langle A_{l'} \rangle
\nonumber\\
&&
{} +  \langle A_i A_l \rangle^c \langle A_{i'} A_j A_{l'} \rangle^c \langle A_{j'}
\rangle
\nonumber\\
&&
+ \langle A_j A_l \rangle^c \langle  A_i A_{j'} A_{l'} \rangle^c \langle A_{i'} \rangle
%\big]
\nonumber\\
&&
{}+\langle A_{i'} A_{j} \rangle^c  \langle A_{j'} A_{l} \rangle^c  \langle A_{i} A_{l'}
\rangle^c
\nonumber\\
&&
{} + \frac 12 \langle A_i A_j A_l \rangle^c  \langle A_{i'} A_{j'} A_{l'} \rangle^c .
\end{eqnarray}

\begin{figure}[t]
\unitlength1cm
\begin{picture}(11,3.2)
\centerline{\psfig{file=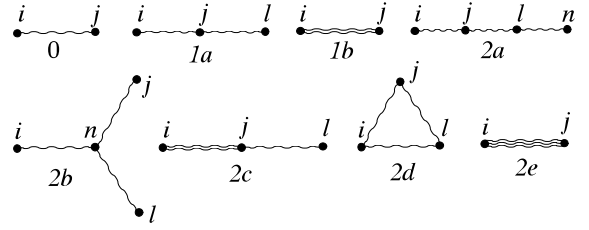,angle=-90,width=17cm}}
\end{picture}
%\par
%
\caption{ \label{cum}
Diagrammatic representation of the cumulant expansion for the quantum
Heisenberg
magnet.
Wavy lines stand for the exchange interaction $J_{ij}$, small solid circles are
cumulants
of spin operators.
}
\end{figure}

The rhs of Eqs.\ (\ref{CumToCum2}) and (\ref{CumToCum3}) are constructed according to
a
principle which can be
formulated
in a diagrammatic language:
The operators $A$ are either contracted into cumulants or connected by the
interaction
lines.
Similarly to other diagram techniques, {\em there are no terms consisting of
disconnected
parts}, e.g., there
is
no term $\langle A_i \rangle \langle A_{i'} \rangle \langle A_j \rangle \langle A_{j'}
\rangle$
in
Eq.\ (\ref{CumToCum2}).
Note that the order of the operators in the cumulants on the rhs of Eqs.\
(\ref{CumToCum2}) and
(\ref{CumToCum3})
is the same as on the left-hand side (lhs).
This fact enables us to write down immediately equations of the form of
Eqs.\ (\ref{CumToCum2}) or (\ref{CumToCum3}), without explicitly performing
steps (i)
and (ii).
The diagrammatic representation of the cumulant expansion of Eq.\ (\ref{TaylorCum}) for the quantum
Heisenberg
magnet in the
zero-field case is
shown in Fig.\ \ref{cum}.
Diagram 0 is the first order of the cumulant expansion in $\hat H$.
It will be shown below that this diagram yields the zeroth order in $1/S$ for the Hamilton function ${\cal H}$, i.e., the classical Hamiltion function  ${\cal H}^{(0)}$.
Diagrams $1a$ and $1b$ correspond to the two terms in Eq.\ (\ref{CumToCum2}).
Diagram $2a$ represent the three first terms in Eq.\ (\ref{CumToCum3}).
These terms have the same topology and differ only by the order of spin
operators.
 This
reflects the
quantum
nature of the latter.
Diagram $2b$ corresponds to the fourth term in Eq.\ (\ref{CumToCum3}).
Diagram $2c$ represent the fifth, sixth, and seventh terms in Eq.\
(\ref{CumToCum3}).
Diagrams $2d$ and $2e$ represent the two last terms, respectively.

\begin{figure}[t]
\unitlength1cm
\begin{picture}(11,3.2)
\centerline{\psfig{file=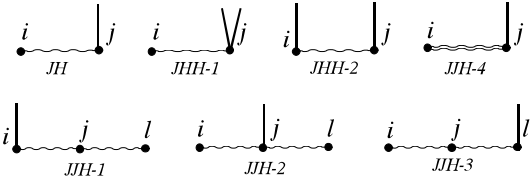,angle=-90,width=17cm}}
\end{picture}
%\par
%
\caption{ \label{cumh}
The same as in Fig.\ \protect\ref{cum} for the mixed field-exchange terms.
Straight lines represent an external magnetic field.
}
\end{figure}

Cumulants corresponding to the mixed field-exchange terms in the
cumulant
expansion can be treated in a similar manner.
One obtains
%\marginpar{CumHJ}
%
\begin{eqnarray}\label{CumHJ}
&&
\frac 12 \langle (A_i A_{i'}) A_j \rangle^c = \langle A_i \rangle  \langle A_{i'} A_j
\rangle^c
\nonumber\\
&&
\frac 12 \langle  A_j (A_i A_{i'}) \rangle^c = \langle A_i \rangle  \langle  A_j A_{i'}
\rangle^c
\nonumber\\
&&
\frac 12 \langle (A_i A_{i'}) A_j A_l\rangle^c =
\langle A_i \rangle  \langle A_{i'} A_j A_l \rangle^c
+ \langle A_i  A_l \rangle^c  \langle A_{i'} A_j \rangle^c
\nonumber\\
&&
\frac{ 1 }{ 2^2 } \langle (A_i A_{i'}) (A_j A_{j'}) A_l \rangle^c =
\langle A_i A_l\rangle^c \langle A_{i'} A_j \rangle^c \langle A_{j'}\rangle
\nonumber\\
&&\qquad
{} + \langle A_i\rangle \langle A_{i'} A_j  A_l\rangle^c \langle A_{j'}\rangle
+ \langle A_i\rangle \langle A_{i'} A_j \rangle^c \langle A_{j'}  A_l \rangle^c
\nonumber\\
&&\qquad\qquad\qquad
{} + \langle A_i A_j A_l \rangle^c \langle A_{i'} A_{j'}\rangle^c,
\end{eqnarray}
etc.
The diagrams corresponding to the rhs of these expressions are shown in
Fig.\
\ref{cumh}.
For the last equation of Eqs.\ (\ref{CumHJ}), in particular, the diagram
corresponding to the $n$th term ($n=1,2,3,4$) on
the rhs is labeled by $JJH$-$n$.

\section{The effective Hamiltonian}
\label{EffHam}

Let us now proceed to the calculation of the different terms in the cumulant
expansion of Eq.\ (\ref{TaylorCum}).
The term of the first order in $\hat H$ yields the classical Hamiltonian
%\marginpar{ClassHam}
%
\begin{equation}\label{ClassHam}
{\cal H}^{(0)} = -  \sum_i {\bf h}_i \cdot {\bf n}_i - \frac 12 \sum_{ij} \tilde J_{ij} {\bf
n}_i
\cdot {\bf
n}_j ,
\end{equation}
which is the zeroth order of the $1/S$ expansion.
Here
%\marginpar{RedVar}
%
\begin{equation}\label{RedVar}
{\bf h}_i \equiv S {\bf H}_i,   \qquad  \tilde J_{ij} \equiv S^2 J_{ij}
\end{equation}
are the reduced magnetic field and the exchange interaction, respectively.
In second and higher orders in $\hat H$ of the cumulant expansion, $1/S$ corrections
appear.
With the help of Eqs.\  (\ref{CumToCum2}), (\ref{CumHJ}), and (\ref{LowestCums})
one
obtains
%\marginpar{HSecond}
%
\begin{eqnarray}\label{HSecond}
&&
{\cal H}^{(1)} =
-\frac \beta S \sum_i ({\bf h}_i \cdot {\bf n}_{i+}) ({\bf n}_{i-} \cdot {\bf h}_i )
\nonumber\\
&&
{} - \frac{ \beta }{ S } \sum_{ij} \tilde J_{ij}
[ ({\bf h}_i \cdot {\bf n}_{i+}) ({\bf n}_{i-} \cdot {\bf n}_j )
+ ({\bf h}_i \cdot {\bf n}_{i-}) ({\bf n}_{i+} \cdot {\bf n}_j ) ]
\nonumber\\
&&
{} -\frac \beta S \sum_{ijl} \tilde J_{ij} \tilde J_{jl}
({\bf n}_i \cdot {\bf n}_{j+}) ({\bf n}_{j-}  \cdot {\bf n}_l )
\nonumber\\
&&
{} - \frac  \beta {S^2} \sum_{ij} \tilde J_{ij}^2
({\bf n}_{i+} \cdot {\bf n}_{j+}) ({\bf n}_{i-}  \cdot {\bf n}_{j-} ) ,
\end{eqnarray}
In this expression, as well as in all other expressions of this type, the transverse
components of the
coherent-state
vectors, ${\bf n}_\pm$, can be expressed in terms of ${\bf n}$ after applying some
vector
algebra
(for details see Appendix).
This is not surprising since ${\bf n}$ is the only vector specifying the spin
coherent state.
Thus one can rewrite  ${\cal H}^{(1)}$  in the form
%\marginpar{HSecondF}
%
\begin{eqnarray}\label{HSecondF}
&&
{\cal H}^{(1)} =
-\frac \beta {4S} \sum_i [{\bf h}_i \times {\bf n}_i]^2
\nonumber\\
&&
{} - \frac \beta {4S} \sum_{ij} \tilde J_{ij}
[({\bf h}_i \cdot {\bf n}_j) - ({\bf h}_i \cdot {\bf n}_i) ({\bf n}_i \cdot {\bf n}_j)
\nonumber\\
&&\qquad\qquad
{} + ({\bf h}_j \cdot {\bf n}_i) - ({\bf h}_j \cdot {\bf n}_j) ({\bf n}_j \cdot {\bf n}_i)]
\nonumber\\
&&
{} -\frac \beta {4S} \sum_{ijl} \tilde J_{ij} \tilde J_{jl}
[ ({\bf n}_i \cdot {\bf n}_l) - ({\bf n}_i \cdot {\bf n}_j) ({\bf n}_j \cdot {\bf n}_l) ]
\nonumber\\
&&
{} - \frac  \beta {16S^2} \sum_{ij} \tilde J_{ij}^2
[1-({\bf n}_i \cdot {\bf n}_j)]^2.
\end{eqnarray}
One can see that the effective classical Hamilton function corresponding to the
Heisenberg
quantum Hamiltonian,
Eq.\ (\ref{QHam}), is of a non-Heisenberg form.
In particular, many-spin interactions appear.

Let us now proceed to the third order terms of the cumulant expansion.
At first we consider the zero-field case.
Equation (\ref{CumToCum3}) generates terms of different topology in ${\cal
H}^{(2)}$
(see Fig.\ \ref{cum})
%\marginpar{HThird}
%
\begin{eqnarray}\label{HThird}
&&
{\cal H}^{(2)} = - \frac{ \beta^2 }{8S^2 } \sum_{ijln} \tilde J_{ij} \tilde J_{jl} \tilde
J_{ln}
\Phi_{2a}( {\bf n}_i, {\bf n}_j, {\bf n}_l,{\bf n}_n)
\nonumber\\
&&\qquad
{} - \frac{ \beta^2 }{12S^2 } \sum_{ijln} \tilde J_{in} \tilde J_{jn} \tilde J_{ln}
\Phi_{2b}( {\bf n}_i, {\bf n}_j, {\bf n}_l,{\bf n}_n)
\nonumber\\
&&\qquad
{} - \frac{ \beta^2 }{24S^3 } \sum_{ijl} \tilde J_{ij}^2 \tilde J_{jl}
\Phi_{2c}( {\bf n}_i, {\bf n}_j, {\bf n}_l)
\nonumber\\
&&\qquad
{} - \frac{ \beta^2 }{48S^3 } \sum_{ijl} \tilde J_{ij} \tilde J_{jl} \tilde J_{li}
\Phi_{2d}( {\bf n}_i, {\bf n}_j, {\bf n}_l)
\nonumber\\
&&\qquad
{} - \frac{ \beta^2 }{48S^4 } \sum_{ij} \tilde J_{ij}^3
\Phi_{2e}( {\bf n}_i, {\bf n}_j) .
\end{eqnarray}
Here the $2a$ term is due to the first, second, and third terms of Eq.\
(\ref{CumToCum3}),
the $2b$ term is due to the fourth term of Eq.\ (\ref{CumToCum3}),
the $2c$ term is due to the fifth, sixth, and seventh terms of Eq.\
(\ref{CumToCum3}),
the $2d$ term is due to the eighth term,
and the $2e$ term is due to the nineth term.
For the functions $\Phi$ in Eq.\ (\ref{HThird}) we find
%\marginpar{Phi3A}
%
\begin{eqnarray}\label{Phi3A}
&&
\Phi_{2a} = \frac{ 16 }{ 3 } \bigg[
({\bf n}_i \cdot {\bf n}_{j+}) ({\bf n}_{j-} \cdot {\bf n}_{l+}) ({\bf n}_{l-} \cdot {\bf
n}_{n})
\nonumber\\
&&\qquad\qquad
{}+({\bf n}_i \cdot {\bf n}_{j+}) ({\bf n}_{j-} \cdot {\bf n}_{l-}) ({\bf n}_{l+} \cdot {\bf
n}_{n})
\nonumber\\
&&\qquad\qquad
{}+({\bf n}_i \cdot {\bf n}_{j-}) ({\bf n}_{j+} \cdot {\bf n}_{l+}) ({\bf n}_{l-} \cdot {\bf
n}_{n})
\bigg]
\nonumber\\
&&\qquad
{} = ({\bf n}_i \cdot {\bf n}_n)
- ({\bf n}_i \cdot {\bf n}_l) ({\bf n}_l \cdot {\bf n}_n)
\nonumber\\
&&\qquad
{}- ({\bf n}_i \cdot {\bf n}_j) ({\bf n}_j \cdot {\bf n}_n)
+ ({\bf n}_i \cdot {\bf n}_j) ({\bf n}_j \cdot {\bf n}_l) ({\bf n}_l \cdot {\bf n}_n)
\nonumber\\
&&\qquad
{}+ \frac 13 \big[
({\bf n}_i \cdot {\bf n}_l) ({\bf n}_j \cdot {\bf n}_n) - ({\bf n}_i \cdot {\bf n}_n) ({\bf
n}_j \cdot
{\bf n}_l)
\big]
\end{eqnarray}
%
%\marginpar{Phi3B}
%
\begin{eqnarray}\label{Phi3B}
&&
\Phi_{2b} = -4 ({\bf n}_i \cdot {\bf n}_{n+}) ({\bf n}_j \cdot {\bf n}_n) ({\bf n}_l \cdot
{\bf
n}_{n-})
\nonumber\\
&&\qquad
{} = ({\bf n}_i \cdot {\bf n}_n) ({\bf n}_j \cdot {\bf n}_n) ({\bf n}_l \cdot {\bf n}_n)
- \frac 13 \bigg[
({\bf n}_i \cdot {\bf n}_n) ({\bf n}_j \cdot {\bf n}_l)
\nonumber\\
&&\qquad
{}+({\bf n}_j \cdot {\bf n}_n) ({\bf n}_i \cdot {\bf n}_l)
+ ({\bf n}_l \cdot {\bf n}_n) ({\bf n}_i \cdot {\bf n}_j)
\bigg],
\end{eqnarray}
%
%\marginpar{Phi3C}
%
\begin{eqnarray}\label{Phi3C}
&&
\Phi_{2c} = -16 \big[
({\bf n}_{i+} \cdot {\bf n}_{j+}) ({\bf n}_{i-} \cdot {\bf n}_j) ({\bf n}_{j-} \cdot {\bf
n}_l)
\nonumber\\
&&\qquad
{}+({\bf n}_{i-} \cdot {\bf n}_{j-}) ({\bf n}_{i+} \cdot {\bf n}_j) ({\bf n}_{j+} \cdot {\bf
n}_l)
\nonumber\\
&&\qquad
{}+({\bf n}_{i+} \cdot {\bf n}_{j+}) ({\bf n}_{i-} \cdot {\bf n}_{j-}) ({\bf n}_j \cdot {\bf
n}_l)
\big]
\nonumber\\
&&
{} = [1-({\bf n}_i \cdot {\bf n}_j)] \big\{
[3({\bf n}_i \cdot {\bf n}_j) - 1] ({\bf n}_i \cdot {\bf n}_l) - 2 ({\bf n}_j \cdot {\bf
n}_l)\big\},
\end{eqnarray}
%
%\marginpar{Phi3D}
%
\begin{eqnarray}\label{Phi3D}
&&
\Phi_{2d} = 4^3 ({\bf n}_{i+} \cdot {\bf n}_{j+}) ({\bf n}_{j-} \cdot {\bf n}_{l+})
({\bf n}_{l-} \cdot {\bf n}_{i-})
\nonumber\\
&&\qquad
{} = ({\bf n}_i \cdot {\bf n}_j)^2 + ({\bf n}_j \cdot {\bf n}_l)^2 + ({\bf n}_l \cdot {\bf
n}_i)^2
\nonumber\\
&&\qquad
{}- ({\bf n}_i \cdot {\bf n}_j) ({\bf n}_j \cdot {\bf n}_l) ({\bf n}_l \cdot {\bf n}_i)
\nonumber\\
&&
{}- \frac 13 \bigg[
({\bf n}_i \cdot {\bf n}_j) + ({\bf n}_j \cdot {\bf n}_l) + ({\bf n}_l \cdot {\bf n}_i)
+ ({\bf n}_i \cdot {\bf n}_j) ({\bf n}_j \cdot {\bf n}_l)
\nonumber\\
&&\qquad
{}+ ({\bf n}_j \cdot {\bf n}_l) ({\bf n}_l \cdot {\bf n}_i)
+ ({\bf n}_l \cdot {\bf n}_i) ({\bf n}_i \cdot {\bf n}_j) \bigg],
\end{eqnarray}
and
%\marginpar{Phi3E}
%
\begin{eqnarray}\label{Phi3E}
&&
\Phi_{2e} = 16 ({\bf n}_{i+} \cdot {\bf n}_{j+}) ({\bf n}_i \cdot {\bf n}_j) ({\bf n}_{i-}
\cdot {\bf
n}_{j-})
\nonumber\\
&&
{} = ({\bf n}_i \cdot {\bf n}_j) [1-({\bf n}_i \cdot {\bf n}_j)]^2.
\end{eqnarray}
The coefficient $\Phi_{3d}$ can be rewritten in the form
%\marginpar{Phi3D1}
%
\begin{equation}\label{Phi3D1}
\Phi_{2d}=-[{\bf n}_i {\bf n}_j {\bf n}_l]^2 + [1-( {\bf n}_i \cdot {\bf n}_j ) ]
[ 1-( {\bf n}_j \cdot {\bf n}_l ) ( {\bf n}_l \cdot {\bf n}_i ) ],
\end{equation}
where $[{\bf a} {\bf b} {\bf c}] \equiv {\bf a} \cdot ({\bf b} \times {\bf c})$ is the
mixed
product and an
appropriate symmetryzation is implied.

The field-dependent terms of order $H^3$, $JH^2$ and, $J^2H$ in ${\cal H}^{(2)}$
can be
calculated with the help of Eq.\ (\ref{CumHJ}).
The result has the form
%\marginpar{HThirdH}
%
\begin{eqnarray}\label{HThirdH}
&&
{\cal H}^{(2)}_{h} = \frac{ \beta^2 }{ 12S^2 }\sum_i ({\bf n}_i \cdot {\bf h}_i)
[{\bf n}_i \times {\bf h}_i]^2
\nonumber\\
&&\qquad
{} - \frac{ \beta^2 }{ 8S^2 }\sum_{ij} \tilde J_{ij}
\Phi_{2h^2}({\bf n}_i, {\bf n}_j, {\bf h}_i, {\bf h}_j)
\nonumber\\
&&\qquad
{} - \frac{ \beta^2 }{ 8S^2 }\sum_{ijl} \tilde J_{ij} \tilde J_{jl}
\Phi_{2h,a}({\bf n}_i, {\bf n}_j, {\bf n}_l,{\bf h}_i, {\bf h}_j, {\bf h}_l)
\nonumber\\
&&\qquad
{} - \frac{ \beta^2 }{ 48S^3 }\sum_{ij} \tilde J_{ij}^2
\Phi_{2h,b}({\bf n}_i, {\bf n}_j, {\bf h}_i, {\bf h}_j),
\end{eqnarray}
where
%\marginpar{Phi3h2}
%
\begin{eqnarray}\label{Phi3h2}
&&
\Phi_{2h^2} = ( {\bf n}_i \cdot {\bf n}_j )
\big[({\bf h}_i \cdot {\bf n}_i)^2 + ({\bf h}_j \cdot {\bf n}_j)^2\big]
\nonumber\\
&&\qquad
{} - \frac 23 \big[({\bf h}_i \cdot {\bf n}_i) ({\bf h}_i \cdot {\bf n}_j)
 + ({\bf h}_j \cdot {\bf n}_j) ({\bf h}_j \cdot {\bf n}_i) \big]
\nonumber\\
&&\qquad
{} - \frac 13 ( {\bf n}_i \cdot {\bf n}_j )[h_i^2 + h_j^2]
\nonumber\\
&&\qquad
{} + ( {\bf h}_i \cdot {\bf h}_j )
-( {\bf h}_i \cdot {\bf n}_i ) ( {\bf n}_i \cdot {\bf h}_j )
-( {\bf h}_i \cdot {\bf n}_j ) ( {\bf n}_j \cdot {\bf h}_j )
\nonumber\\
&&\qquad
{}
+ ( {\bf h}_i \cdot {\bf n}_i ) ( {\bf n}_i \cdot {\bf n}_j ) ( {\bf n}_j \cdot {\bf h}_j )
\nonumber\\
&&\qquad
{} + \frac 13 \big[ ( {\bf h}_i \cdot {\bf n}_j ) ( {\bf h}_j \cdot {\bf n}_i )
- ( {\bf h}_i \cdot {\bf h}_j ) ( {\bf n}_i \cdot {\bf n}_j ) \big],
\end{eqnarray}
%
%\marginpar{Phi3ha}
%
\begin{eqnarray}\label{Phi3ha}
&&
\Phi_{2h,a} = ( {\bf h}_i \cdot {\bf n}_l ) -
({\bf h}_i \cdot {\bf n}_j ) ({\bf n}_j \cdot {\bf n}_l )
\nonumber\\
&&\qquad
{} + ( {\bf h}_l \cdot {\bf n}_i ) -
({\bf h}_l \cdot {\bf n}_j ) ({\bf n}_j \cdot {\bf n}_i )
\nonumber\\
&&\qquad
{} - \big[ ( {\bf h}_i \cdot {\bf n}_i ) + ( {\bf h}_l \cdot {\bf n}_l ) \big]
\big[ ({\bf n}_i \cdot {\bf n}_l) -
({\bf n}_i \cdot {\bf n}_j) ({\bf n}_j \cdot {\bf n}_l)\big]
\nonumber\\
&&\qquad
{} + \frac 13 \bigg[ ({\bf n}_i\cdot{\bf n}_l)
\big[({\bf h}_i \cdot {\bf n}_j) + ({\bf h}_l \cdot {\bf n}_j)\big]
\nonumber\\
&&\qquad
{}- ({\bf h}_i \cdot {\bf n}_l) ({\bf n}_i\cdot{\bf n}_j)
- ({\bf h}_l \cdot {\bf n}_i) ({\bf n}_l\cdot{\bf n}_j) \bigg]
\nonumber\\
&&\qquad
{} + 2({\bf n}_i\cdot{\bf n}_j)({\bf h}_j\cdot{\bf n}_j)({\bf n}_j\cdot{\bf n}_l)
- \frac 23 \bigg[ ({\bf n}_i\cdot{\bf n}_l)({\bf h}_j\cdot{\bf n}_j)
\nonumber\\
&&\qquad
{}
+ ({\bf n}_j\cdot{\bf n}_l)({\bf h}_j \cdot {\bf n}_i )
+ ({\bf n}_j\cdot{\bf n}_i)({\bf h}_j \cdot {\bf n}_l ) \bigg],
\end{eqnarray}
and
%\marginpar{Phi3hb}
%
\begin{eqnarray}\label{Phi3hb}
&&
\Phi_{2h,b} = -[1-({\bf n}_i\cdot{\bf n}_j)]
\big\{
2[({\bf h}_i\cdot{\bf n}_j)+({\bf h}_j\cdot{\bf n}_i)]
\nonumber\\
&&\qquad
{} +
[1-3({\bf n}_i\cdot{\bf n}_j)][({\bf h}_i\cdot{\bf n}_i)+({\bf h}_j\cdot{\bf n}_j)]
\big\}.
\end{eqnarray}
For a homogeneous field, ${\bf h}_i = {\bf h}$, the last expression becomes
%\marginpar{Phi3hbHom}
%
\begin{equation}\label{Phi3hbHom}
\Phi_{2h,b} = -3[{\bf h} \cdot ({\bf n}_i + {\bf n}_j)][1-({\bf n}_i\cdot{\bf n}_j)]^2 .
\end{equation}

As was said after Eq.\ (\ref{TaylorCum}), cumulant expansion is, in general, an
expansion
in $\beta/S$.
In particular, in zero field in the $n$th order of the cumulant expansion, terms of
order
$\tilde J(\beta\tilde J/S)^{n-1}$ appear.
On the other hand, there are also terms carrying additional powers of $1/S$, such
as the
last term of Eq.\
(\ref{HSecondF})
and the three last terms of Eq.\ (\ref{HThird}).
This feature results, formally, from reexpressing of cumulants of composite operators
by
ordinary cumulants,
Eqs.\  (\ref{CumToCum2}) and (\ref{CumToCum3}), which is specific to many-spin
systems.
For $S\gg 1$, the terms containing powers of $\beta\tilde J/S$ without additional factors $1/S$ dominate over
the
other terms in the
whole range of
temperatures.
Those terms are given by ``tree'' diagrams, such as 0, $1a$, $2a$, and $2b$ in
Fig.\
\ref{cum}.
These diagrams are maximally branched and they do not contain loops or parallel
interaction lines.
Non-tree diagrams can be obtained from the tree diagrams by joining two solid
circles
into one.
This leads each time to appearance of a spin cumulant of a higher order and
thus to
an additional factor
of $1/S$
[see Eq.\  (\ref{LowestCums})].
Minimally branched diagrams are those consisting of several parallel wavy lines,
such as
the diagrams $1b$ and $2e$ in Fig.\ \ref{cum}.
These diagrams result in terms of order $\tilde J(\beta\tilde J/S^2)^{n-1}$ in
${\cal
H}$.

\section{Physical quantities}
\label{observables}

Quantum-statistical averages of operators describing various physical quantities
can be
obtained by differentiation of the partition function ${\cal Z}$ or its logarithm with respect to
appropriate
parameters.
For the internal energy $U = \langle \langle \hat H \rangle \rangle$ one obtains
%\marginpar{UHStar}
%
\begin{equation}\label{UHStar}
U = - \partial \ln {\cal Z} / \partial \beta  = \langle {\cal H}^*\rangle,
\end{equation}
where $\langle \ldots \rangle$ denotes a classical thermal average and
%\marginpar{HStar}
%
\begin{equation}\label{HStar}
{\cal H}^* = \partial (\beta{\cal H})/\partial \beta
= {\cal H}^{(0)} + 2 {\cal H}^{(1)} + 3 {\cal H}^{(2)} + \ldots
\end{equation}
differs from ${\cal H} = {\cal H}^{(0)} + {\cal H}^{(1)} + {\cal H}^{(2)} + \ldots$
The scaled magnetization per site,
${\bf m}\equiv \langle \langle{\bf S} \rangle\rangle/(SN) \equiv \sum_i
\langle\langle {\bf S}_i \rangle\rangle/(SN)$,
is given by
%\marginpar{mnStar}
%
\begin{equation}\label{mnStar}
{\bf m} = \frac 1N \frac{ \partial \ln {\cal Z} }{ \partial (\beta {\bf h}) }
= \frac 1N \langle {\bf n}^*\rangle,
\qquad
{\bf n}^* = - \frac{\partial {\cal H} }{\partial {\bf h} },
\end{equation}
where ${\bf h}$ is the homogeneous part of the magnetic field defined formally by
${\bf h}_i \Rightarrow {\bf h}_i + {\bf h}$.
Here, ${\bf n}^*$ is not just ${\bf n} = \sum_i {\bf n}_i$ but contains quantum
corrections from all orders of the cumulant expansion.
In particular, to first order of the cumulant expansion, one obtains from
Eqs.\ (\ref{ClassHam}) and (\ref{HSecondF})
%\marginpar{nStar}
%
\begin{eqnarray}\label{nStar}
&&
{\bf n}^* = \sum_i {\bf n}_i +
\frac \beta{2S} \sum_i [ {\bf h}_i - {\bf n}_i ({\bf n}_i \cdot {\bf h}_i)]
\nonumber\\
&&\qquad
{} + \frac \beta{4S} \sum_{ij}  \tilde J_{ij}
[1-({\bf n}_i \cdot {\bf n}_j)] ({\bf n}_i + {\bf n}_j).
\end{eqnarray}

The reduced correlation function of different spin components on different lattice
sites
can be written as
%\marginpar{CF}
%
\begin{eqnarray}\label{CF}
&&
\frac 1{S^2} \langle \langle S_{i\alpha} S_{j\beta} \rangle \rangle =
\frac 1{\cal Z} \frac{\partial^2 {\cal Z} }
{ \partial (\beta h_{i\alpha}) \partial (\beta h_{j\beta}) }
\nonumber\\
&&\qquad
{} = \left\langle \frac{ \partial {\cal H} }{ \partial h_{i\alpha} }
\frac{ \partial {\cal H} }{ \partial h_{j\beta} }
- \frac{ \partial^2 {\cal H} }
{ \partial h_{i\alpha} \partial h_{j\beta} } \right\rangle.
\end{eqnarray}
One notices that in order to calculate a correlation function, it is insufficient to perform a
classical thermal average of $n_{i\alpha} n_{j\beta}$ or even of $n_{i\alpha}^*
n_{j\beta}^*$.
The last term of Eq.\ (\ref{CF}) makes a contribution to third order in the
cumulant expansion due to the terms of the type $JH^2$ [see Eq.\
(\ref{HThirdH})].

\section{Application to the spin chain}
\label{chain}

The isotropic spin chain in zero magnetic field is a simply solvable model in the classical limit
\cite{fis64amjp}.
The effective quasiclassical Hamiltonian discussed in this paper can be used to
analytically calculate $1/S$ corrections to the classical results.
The $1/S$ expansion of the partition function has the form
%\marginpar{ZExpGen}
%
\begin{eqnarray}\label{ZExpGen}
&&
{\cal Z} \cong \tilde{\cal  Z}_0 \left[ 1 - \langle\beta( {\cal H}^{(1)} +  {\cal H}^{(2)} )
\rangle
+ \frac 1 {2!}  \langle [\beta {\cal H}^{(1)}]^2 \rangle +\ldots \right],
\end{eqnarray}
where the averages are performed with respect to the classical Hamiltonian ${\cal
H}^{(0)}$,
%\marginpar{tilZ0}
%
\begin{equation}\label{tilZ0}
\tilde {\cal Z}_0 = \left( \frac {2S+1} {4\pi} \right)^N  {\cal Z}_0,
\end{equation}
and ${\cal Z}_0$ is the partition function of the classical system.
For the open spin chain the latter is given by
${\cal Z}_0 = 4\pi  ( 4\pi  \sinh (\xi) /\xi )^{N-1}$ with $\xi \equiv \beta \tilde J$.
To order $1/S^2$, one should use for the linear ${\cal H}^{(1)}$ term in Eq.\ (\ref{ZExpGen}) the third and fourth terms of
Eq.\ (\ref{HSecondF}),  for the quadratic ${\cal H}^{(1)}$ term the third term  of Eq.\
(\ref{HSecondF}),  and for the linear ${\cal H}^{(2)}$ term the first and second terms
of Eq.\ (\ref{HThird}).
After performing thermodynamic averages one obtains
%\marginpar{LnZChain}
%
\begin{eqnarray}\label{LnZChain}
&&
\frac {\ln {\cal Z}} N \cong \ln(2S+1) + \ln \left( \frac{ \sinh \xi }{ \xi } \right)+ \frac{ \xi B
}{S}
\nonumber \\
&&
\qquad{}
+ \frac{ 5\xi^2 -7\xi^2B^2 \mp \xi^2 B -9\xi B}{ 12S^2 }
+ O\left(\frac{ 1}{ S^3 }\right),
\end{eqnarray}
for ferro- and antiferromagnets, where $B \equiv \coth \xi - 1/\xi$ is the Langevin
function.
This formula was obtained earlier \cite{ulyzas92}
with the help of the Wigner-Kirkwood expansion which avoids using effective
classical Hamiltonians.
It strongly resembles the result for the two-spin model, Eq.\ (20) of Ref.\
\cite{klafulgar99epl}, where the factor 7 is replaced by 6.
For the energy per spin $U = - \partial \ln {\cal Z} / \partial (\beta N) $ one obtains
%\marginpar{UChain}
%
\begin{eqnarray}\label{UChain}
&&
\frac U {\tilde J} \cong  -B - \frac 1S [ B + \xi B' ] - \frac 1{S^2} [ 10\xi
\nonumber \\
&&
\qquad{}- (14\xi B + 9)(B + \xi B' )
 \mp (2B + \xi B' )].
\end{eqnarray}

 Let us consider now the energy  $U$ per spin
which follows from  linear spin-wave theory
%\marginpar{USWT}
%
\begin{eqnarray}\label{USWT}
&&
U = - \frac{\tilde J_0}2 - \frac{\tilde J_0}{2S}
\int\!\!\! \frac{ d {\bf q}}{ (2\pi)^d } (1-\tilde\varepsilon_{\bf q})
+T\!\!\int\!\!\! \frac{ d {\bf q}}{ (2\pi)^d }
\frac{ \tilde\beta \tilde\varepsilon_{\bf q} }{\exp(\tilde\beta\tilde\varepsilon_{\bf q})
-1}
\nonumber\\
&&
\tilde\beta \equiv \beta \tilde J_0/S, \qquad\!\!
\tilde\varepsilon_{\bf q} \equiv (S/\tilde J_0)\varepsilon_{\bf q}
= \left\{
\begin{array}{ll}
\displaystyle
1-\lambda_{\bf q},           &  {\rm F}\\
\displaystyle
\sqrt{1-\lambda_{\bf q}^2},        & {\rm AF}.
\end{array}
\right.
\end{eqnarray}
Here $\tilde J_0$ is the zeroth Fourier component of $\tilde J_{ij}$ and
$\lambda_{\bf q}\equiv \tilde J_{\bf q}/\tilde J_0$.
The first term in the expression for $U$ is the classical ground-state energy, the
second
term is the quantum correction to the former, and the last term is the
temperature-dependent magnon contribution.
Strictly speaking, linear spin-wave theory is only applicable for higher then two dimensions, but,
although, Eq.\ (\ref{USWT}) remains well-defined in lower dimensions.
The Haldane gap in the magnon spectrum for integer $S$, which is not taken into
account in Eq.\ (\ref{USWT}), behaves as $\exp(-S)$ and becomes negligible for large
spins.
Whereas $\tilde\varepsilon_{\bf q}$ is of order unity, the parameter $\tilde\beta$ is
precisely the small parameter of the cumulant expansion.
If one expands $U$ in powers of  $\tilde\beta$, one obtains a series which is very
close to the one following from the cumulant expansion in the limit $\xi\equiv \tilde J/T
\gg 1$.
In particular, for a ferromagnetic chain in the temperature interval
$\tilde J/S \ll T \ll \tilde J$  Eq.\ (\ref{UChain}) yields
%\marginpar{UOverlap}
%
\begin{equation}\label{UOverlap}
\frac U {\tilde J} \cong - 1 - \frac 1 S + \frac 1 \xi + \frac \xi {2S^2} - \frac 1 {2S^2} ,
\end{equation}
whereas expanding Eq.\ (\ref{USWT}) we obtain the same expression without the last
term.
In the antiferromagnetic case one obtains similar expressions with coefficients 6 and 3
instead of 2 and 2 in the denominator.

\begin{figure}[t]
\unitlength1cm
\begin{picture}(11,6)
\centerline{\psfig{file=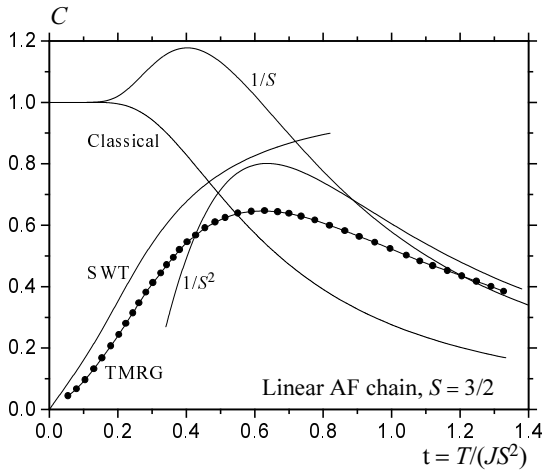,angle=-90,width=8cm}}
\end{picture}
\caption{ \label{cum_c}
Heat capacity of the antiferromagnetic Heisenberg $S=3/2$ chain.
Theoretical results to zeroth, first and second order of the cumulant expansion, as well
as of linear spin-wave theory.
They  are compared with the numerical TMRG calculation
of Ref.\  \protect\cite{xia98}.
}
\end{figure}

In Fig.\ \ref{cum_c} we compare the heat capacity $C=dU/dT$ of a Heisenberg
antiferromagnetic chain with $S=3/2$ calculated in different orders of the cumulant expansion [see
Eq.\ (\ref{UChain})] and following from the linear spin-wave theory [see Eq.\
(\ref{USWT})] with numerical result of Ref.\ \cite{xia98} where the
transfer-matrix renormalization group (TMRG) was used.
Although a spin value 3/2 is not large, one notices that taking into account
quantum corrections improves the behavior of the heat capacity provided the
temperature is not too low.
In order to achieve quantitative agreement with the numerically accurate result, one must take into
account higher-order corrections in $1/S$.
This requires calculation of the next terms of the effective classical Hamiltonian, which
can be done with the help of the diagram technique developed in Sec.\ \ref{cumexp}.
Also on the low-temperature side the accuracy of the SWT result can be improved by
taking in Eq.\ (\ref{USWT}) into account $1/S$ corrections to the magnon spectrum.

\section{Discussion}
\label{discussion}

In the preceding part of the paper, we have obtained an effective classical
Hamiltonian
${\cal H}$ for the large-spin quantum magnet described by the Heisenberg
Hamiltonian
$\hat H$, Eq.\ (\ref{QHam}).
${\cal H}$ consists of a purely classical part and quantum
corrections of
different orders in $1/S$.
Quantum corrections have a non-Heisenberg form and their structure becomes
more and
more
complicated with increasing order of $1/S$.
In particular, pair interactions in $\hat H$ give rise to many-spin interactions in
${\cal H}$.

Our effective classical Hamilton function ${\cal H}$ looks different from the effective
classical
Hamilton function
obtained in Refs.\ \cite{cuctogvaiver969798}.
The latter has the same Heisenberg form, and the quantum effects are absorbed
in the
factor renormalizing the exchange interaction.
This factor satisfies a system of nonlinear equations.
It is difficult to make a direct comparison of the two effective classical
Hamilton functions
because of their different structures and different ways of derivation: Our approach is perturbative in $1/S$ and it leads to the same results for the physical quantities as the Wigner-Kirkwood expansion, whereas the approach of Refs.\ \cite{cuctogvaiver969798} is nonperturbative.

The effective classical Hamilton function ${\cal H}$ obtained in this paper can be used
to
compute quantum corrections for magnetic systems with large spins and not too
low
temperatures.
For one-dimensional models, this was done with the help of analytical
methods.
In two and higher dimensions, one can apply the diagram technique for classical
spins
(see, e.g., Ref.\ \cite{gar96prb}), which should be generalized to
non-Heisenberg Hamiltonians, however.
For models without long-range order, such as two-dimensional ferro- and
antiferromagnets, it is problematic to sum up the relevant diagrams for the
corresponding classical case when the temperature is low.
Then the $1/D$ expansion, where $D$ is the number of spin components, proves
to be an
efficient tool for low-dimensional classical magnets
\cite{gar94jsp,gar96jsp,hinnowgar99prl}.
Since ${\cal H}$ is written in terms of various scalar products, it can be easily
generalized to arbitrary values of $D$ and treated with the help of a $1/D$
expansion.

\section*{Appendix: Elimination of transverse components of the
coherent-state
vectors}

There are two generic formulas for the elimination of transverse components of
the
coherent-state vectors.
The first one,
%\marginpar{ab1}
%
\begin{equation}\label{ab1}
\sum_{\alpha=x,y} ({\bf a} \cdot {\bf n}_\alpha ) ({\bf n}_\alpha \cdot {\bf b} )
= ( {\bf a} \cdot {\bf b} ) - ({\bf a} \cdot {\bf n} ) ({\bf n} \cdot {\bf b} ),
\end{equation}
where ${\bf a}$ and ${\bf b}$ are arbitrary vectors, follows from the definition of
the scalar
product
$({\bf a} \cdot {\bf b})$.
The second formula is
%\marginpar{ab2}
%
\begin{eqnarray}\label{ab2}
&&
({\bf a} \cdot {\bf n}_x ) ({\bf n}_y \cdot {\bf b} ) - ({\bf a} \cdot {\bf n}_y ) ({\bf n}_x
\cdot
{\bf b} ) =
([{\bf n}_x \times {\bf n}_y] \cdot [{\bf a} \times {\bf b} ] )
\nonumber\\
&&\qquad\qquad\qquad\qquad
 = ( {\bf n} \cdot [{\bf a} \times {\bf b} ]).
\end{eqnarray}
Combining these two formulas one obtains the relation
%\marginpar{ab3}
%
\begin{eqnarray}\label{ab3}
&&
4({\bf a} \cdot {\bf n}_\pm ) ({\bf n}_\mp \cdot {\bf b} ) =
( {\bf a} \cdot {\bf b} ) - ({\bf a} \cdot {\bf n} ) ({\bf n} \cdot {\bf b} )
\nonumber\\
&&\qquad\qquad\qquad
{}\pm i ( {\bf n} \cdot [{\bf a} \times {\bf b} ]),
\end{eqnarray}
which is used in the main text to eliminate ${\bf n}_\pm$.
Other useful relations are
%\marginpar{npmRels}
%
\begin{eqnarray}\label{npmRels}
&&
({\bf n}_+ \cdot {\bf n}_-) = \frac 14 (n_x^2 + n_y^2) = \frac 12
\nonumber\\
&&
[{\bf n}_+ \times {\bf n}_-] = \frac i2 {\bf n},
\qquad
[{\bf n}_\pm \times {\bf n}] = \mp i {\bf n}_\pm.
\end{eqnarray}
%

%\bibliography{gar}

\end{document}